\documentclass[preprintnumbers,amsmath,amssymb]{revtex4}
\usepackage{graphics}
\usepackage{dcolumn}

\begin{document}
\title{Analytic average magnetization expression for the body centered cubic Ising lattice}
\author{Tuncer Kaya}
\address{ Department of Physics, Yildiz Technical University,
        34220 Davutpa\c sa, Istanbul, Turkey}

\begin{abstract}
Recently we have performed large-scale average magnetization
studies of the Ising model for the square, honeycomb, triangular,
and simple cubic lattice. We want to complement those studies with
the structurally more complicated body-centered cubic lattice
Ising model in this paper. We have relevantly calculated the order
parameter or average magnetization expression for the
body-centered cubic Ising lattice based on the recently developed
interrelation, given by $\langle\sigma_{0, i}\rangle=
\langle\tanh[K(\sigma_{1,i}+\sigma_{2,i}+\dots
+\sigma_{z,i})+H]\rangle$. We will use the same conjectures for
the odd spins correlation functions as in our previous works. The
mathematical forms of the odd spin correlation functions will be
eventually elucidated by some physical discussions. We have seen
that the obtained average magnetization expression with the
numerical data for the body-centered cubic lattice Ising model is
in good agreement.
\end{abstract}

\maketitle
\section{Introduction}
The Ising model was originated by Lenz in 1920. Later, it was
investigated by his student Ising in 1925 \cite{Ising1}. The model
is very important in the investigation of phase transition since
it unifies the study of phase transitions in systems as diverse as
gas--liquids, ferromagnets, binary alloys, and so
on\cite{Stanley1,Hu1,Zhu1,Keskin1,Aouini1}. The model has an exact
solution in one-dimension (1D) with the prediction of the absence
of phase transition at finite temperature
\cite{Huang1,Baxter1,book1,Beale1,McCOY1}. The model has been
studied using several theoretical approaches, such as high and
low-temperature series expansions, Monte Carlo simulations,
renormalization group methods, and perturbative field theoretical
methods
\cite{Hasenbusch1,Jasch1,Gupta1,Heringa1,Salman1,Campostrini1,Butera1}.
A more extensive list of methods and references can be found in
the exhaustive review article by Pelissetto and Vicari
\cite{Pelissetto1}. Almost seventy years have passed since
Onsager's \cite{Onsager1} celebrated solution of the Ising model
free energy in 1944, followed by Yang's \cite{Yang1} proof of
Onsager's result for the spontaneous magnetization in 1952, there
is, however, no known solution for the 3D model
\cite{Zhang1,Wu41,Perk51}. Therefore, Onsager and Yang's works of
the square lattice Ising model is still one of the very few
exactly solved models \cite{Yang831,Perk11,Perk21}, it now serves
as a proving ground for new theories, approximations, and
numerical algorithms. Before introducing the calculation of the
average magnetization expressions for the body-centered cubic
lattice, we think that it is relevant to mention shortly the
important achievements of the subject.

The proof of the existence of the critical point by Kramers and
Wannier \cite{Kramers1} with the method of a dual transformation
was an important development in the investigation of phase
transition. Later, the qualitative calculation of Kramers and
Wannier was proven analytically by Onsager and \cite{Newell1,
Kaufman1}. The spontaneous magnetization of the Ising model on
rectangular lattice was first calculated by Onsager, he has not,
however, published his derivation \cite{Keh1}. The average
magnetization expression of the square lattice Ising model,
$\langle\sigma\rangle=[1-\sinh(2K)^{-4}]^{\frac{1}{8}}$ was first
published by Yang \cite{Yang19521}. The method used in Yang's
derivation is too cumbersome, he recalls it as the longest
calculation in his career \cite{201}. Later, a less complicated
method, but is still too hard to recover, has been proposed to
study the 2D Ising model \cite{261,271,281,331,341, Syozi1}. It is
also important to mention that average magnetization relations
were obtained for the Ising model on honeycomb lattice
\cite{Naya1} by Naya and by \cite{Potts1} for the triangular
lattice. Now let us introduce the main steps that will be used in
the calculation of the average magnetization of the body-centered
cubic lattice.  For the sake of confirmation of the validity and
relevance of the method that will be used in this paper, we first
want to mention some important features of the odd spins
correlation function since they will be one important part of the
calculation of the average magnetization of body-centered cubic
Ising lattice in his paper. Unfortunately, there is no large
available literature except for some papers on three spin
correlation functions. The three spin correlation function of the
2D Ising model \cite{1141,Baxtery1} was first considered by Baxter
for three spins surrounding a triangle by using the Pfaffian
method. A simpler derivation of the three spin correlation of the
honeycomb lattice was also given by Enting \cite{Enting1}. We
think that it is important to mention that there were some other
important studies on the subject of the three spin correlation
functions \cite{1041,1051,1061,Tanaka1}. The common physical
properties of the three spin correlation function obtained in
these works is that: the three spin correlation function
manifestly assumes the same critical coupling strength and the
same critical exponent as the order parameter. As we will see in
the next section, these physical properties are relevant as well
as necessary to describe the three spin correlation function. We
can, therefore, use these physical properties to propose a
relevant mathematical functional form for the odd spin correlation
functions. These points will be elucidated more through this
paper. In the next section, we will try to obtain the desired
average magnetization expression for the body-centered cubic
lattice with the same analytic method used in our previous paper
\cite{kaya21,kaya31} and the previously derived average
magnetization interrelation by us \cite{kaya1}. The obtained
expression will be compared to the already available simulation
result \cite{Hakan1}. In the same section, we will discuss the
relevance of the obtained average magnetization relation and we
will also present some concluding remarks about the obtained
relation.

\section{The average magnetization calculation of body-centered cubic Ising lattice}

The previously derived formula \cite{kaya1} can be arranged in the absence of external magnetic field for the
body-centered cubic lattice as,
\begin{equation}
\langle\sigma_{0,i}\rangle=
\langle\tanh[ K(\sigma_{1,i}+\sigma_{2,i}+\cdots\sigma_{8,i} )]\rangle.
\end{equation}
Here $\sigma_{0,i}$ denotes the
central spin at the $i^{th}$ site while $\sigma_{l,i}$,
$l=1,2,\cdots,8$, are the nearest neighbor spins around its central
spin,and $K$ is the coupling strength. Apparently from now on, caring the index $i$ is not necessary.

Expressing the tangent hyperbolic function with the following equivalent mathematical form by the help of Fig. 1 as,
\begin{eqnarray*}
&&\tanh[ K(\sigma_{1}+\sigma_{2}+\cdots +\sigma_{8})]=A_{1}\{\sigma_{1}+\sigma_{2}+\cdots+\sigma_{8} \}+
{} \\ && A_{2}\{\sigma_{1}\sigma_{2}\sigma_{3}+\cdots\}+A_{3}\{\sigma_{1}\sigma_{2}\sigma_{3}\sigma_{4}\sigma_{5}+\cdots\}
+A_{4}\{\sigma_{1}\sigma_{2}\sigma_{3}\sigma_{4}\sigma_{5}\sigma_{6}\sigma_{7}+\cdots\}.
\end{eqnarray*}

It is worthwhile to mention that in the parentheses of this equation indicated as $\{\cdots\}$ coming after
$A_{1}$, $A_{2}$, $A_{3}$, and $A_{4}$, there are $8, 56, 56, 8$ terms respectively.
\begin{figure*}[!hbt]
\begin{center}$
\begin{array}{cc}
\scalebox{0.7}{\includegraphics{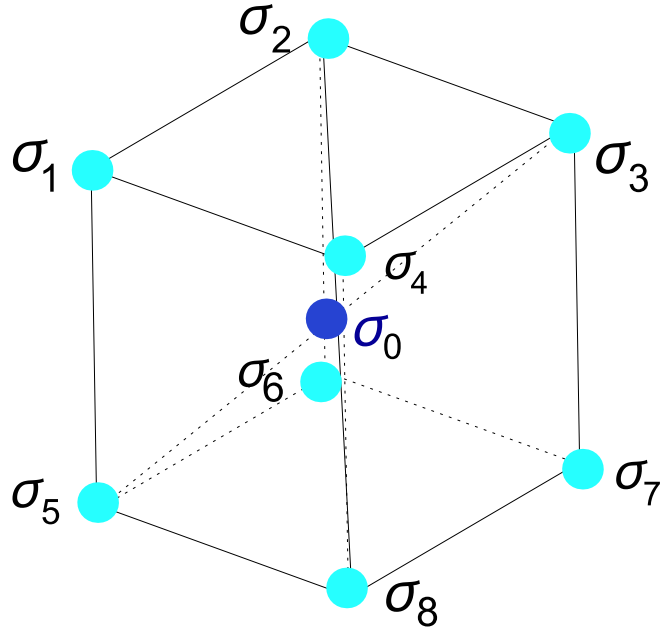}}
\end{array}$
\end{center}
\caption{The body centered cubic lattice structure, the eight spins surrounding the central spin $\sigma_{0}.$}
\end{figure*}

Writing the relation $\tanh[ K(\sigma_{1}\!+\!\sigma_{2}\!+\cdots \!+\sigma_{8})]$ for different orientations of
the eight spin degrees of freedom, one can only obtain four linearly independent equation as
\begin{eqnarray*}
&&\tanh(8K)=8A_{1}+56A_{2}+56A_{3}+8A_{4},
{} \nonumber \\ && \tanh(6K)=6A_{1}+14A_{2}-14A_{3}-6A_{4},
{} \nonumber \\ && \tanh(4K)=4A_{1}-4A_{2}-4A_{3}+4A_{4},
{} \nonumber\\ && \tanh(2K)=2A_{1}-6A_{2}+6A_{3}-2A_{4}.
\end{eqnarray*}

The solution of this system of linear equations, after some algebra the solution for $A_{1}$, $A_{2}$, $A_{3}$,
and $A_{4}$, can be obtained readily,
\begin{eqnarray*}
&&A_{1}(K)=\frac{1}{128}[14\tanh(2K)+14\tanh(4K)+6\tanh(6K)+\tanh(8K)],
{} \nonumber \\ &&A_{2}(K)= \frac{1}{128}[-6\tanh(2K)-2\tanh(4K)+2\tanh(6K)+\tanh(8K)],
{} \nonumber \\ && A_{3}(K)=\frac{1}{128}[6\tanh(2K)-2\tanh(4K)-2\tanh(6K)+\tanh(8K)],
{} \nonumber\\ && A_{4}(K)= \frac{1}{128}[-14\tanh(2K)+14\tanh(4K)-6\tanh(6K)+\tanh(8K)],
\end{eqnarray*}
by these expressions.

Now, substituting the equivalent mathematical form of tangent hyperbolic function into Eq. (1) and followingly
taking the average of both sides, Eq. (1) turns out to be
\begin{eqnarray}
&&\langle\sigma\rangle=8A_{1}\langle\sigma\rangle+A_{2}[ 24\langle\sigma_{1}\sigma_{2}\sigma_{3}\rangle+
24\langle\sigma_{1}\sigma_{2}\sigma_{7}\rangle+ 8\langle\sigma_{1}\sigma_{3}\sigma_{6}\rangle]
{} \nonumber \\ && +A_{3}[ 24\langle\sigma_{1}\sigma_{2}\sigma_{3}\sigma_{4}\sigma_{5}\rangle+
24\langle\sigma_{1}\sigma_{2}\sigma_{7}\sigma_{8}\sigma_{3}\rangle+8\langle\sigma_{1}\sigma_{2}\sigma_{3}\sigma_{5}
\sigma_{7}\rangle]{} \nonumber \\ &&+8A_{4}\langle\sigma_{1}\sigma_{2}\sigma_{3}\sigma_{4}\sigma_{5}
\sigma_{6}\sigma_{7}\rangle.
\end{eqnarray}

Now let us introduce some physical interpretation of the odd spins correlation functions appearing in this equation.
To this end, taking into account the critical behavior of the order parameter $\langle\sigma\rangle$ might be helpful.
Recalling the singularity of the order parameter at the critical point, one can easily see that the odd spins
correlation function can be expressed as a function of the order parameter. On the other hand, it is not easy or not
even possible to propose an exact relation for the odd spins correlations, but taking into account general behavior
of the critical properties of the order parameter, one can conjecture or propose heuristic functional forms for the
odd spins correlation functions with the following relations as,
\begin{eqnarray*}
&&\langle\sigma_{1}\sigma_{2}\sigma_{3}\rangle=a_{2,1}\langle\sigma\rangle+(1-a_{2,1})
\langle\sigma\rangle^{\frac{1+\beta}{\beta}},
{} \nonumber \\ &&\langle\sigma_{1}\sigma_{2}\sigma_{7}\rangle=a_{2,2}\langle\sigma\rangle+
(1-a_{2,2})\langle\sigma\rangle^{\frac{1+\beta}{\beta}},
{}\nonumber \\ && \langle\sigma_{1}\sigma_{3}\sigma_{6}\rangle=a_{2,3}\langle\sigma\rangle+
(1-a_{2,3})\langle\sigma\rangle^{\frac{1+\beta}{\beta}},
{} \nonumber \\ && \langle\sigma_{1}\sigma_{2}\sigma_{3}\sigma_{4}\sigma_{5}\rangle=a_{3,1}\langle\sigma\rangle+
(1-a_{3,1})\langle\sigma\rangle^{\frac{1+\beta}{\beta}},
{} \nonumber \\ && \langle\sigma_{1}\sigma_{2}\sigma_{3}\sigma_{7}\sigma_{8}\rangle=a_{3,2}\langle\sigma\rangle+
(1-a_{3,2})\langle\sigma\rangle^{\frac{1+\beta}{\beta}},
{} \nonumber \\ && \langle\sigma_{1}\sigma_{2}\sigma_{3}\sigma_{5}\sigma_{7}\rangle=a_{3,3}\langle\sigma\rangle+
(1-a_{3,3})\langle\sigma\rangle^{\frac{1+\beta}{\beta}},
{} \nonumber \\ && \langle\sigma_{1}\sigma_{2}\sigma_{3}\sigma_{4}\sigma_{5}\sigma_{6}\sigma_{7}\rangle=
a_{4,1}\langle\sigma\rangle+(1-a_{4,1})\langle\sigma\rangle^{\frac{1+\beta}{\beta}}.
\end{eqnarray*}
Substituting these conjectured odd spins correlation functions into Eq. (2), it leads to
\begin{figure*}[!hbt]
\begin{center}$
\begin{array}{cc}
\scalebox{1.2}{\includegraphics{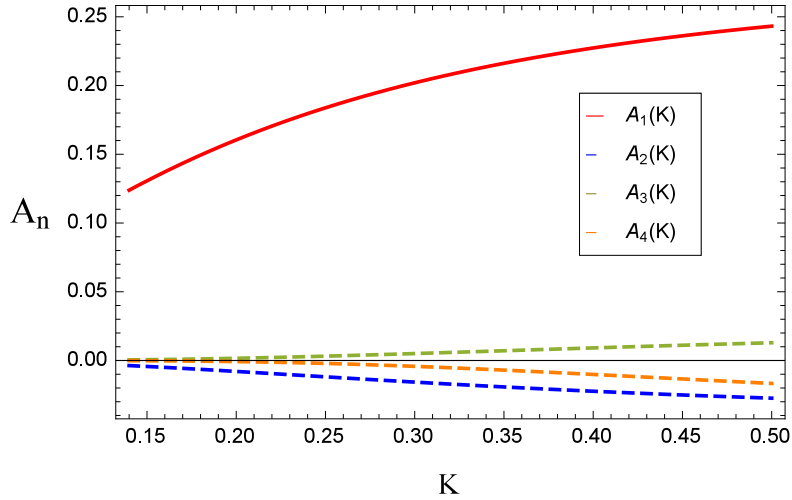}}
\end{array}$
\end{center}
\caption{The plot of $A_{1}$, $A_{2}$, $A_{3}$, and $A_{4}$ with respect to $K$. }
\end{figure*}
\begin{eqnarray*}
&&\!\!\!\!\!\!\!\!\!\!\textstyle\langle\sigma\rangle=\Big\{ 8A_{1}\!+\!24A_{2}\Big[a_{2,1}\!+\!a_{2,2}\!+
\!\frac{a_{2,3}}{3}\Big]\!+\!24A_{3}\Big[a_{3,1}\!+\!a_{3,2}\!+\!\frac{a_{3,3}}{3}\Big]\!+\!8A_{4}a_{4,1}\Big\}
\langle\sigma\rangle {}\nonumber \\ &&\textstyle+\Big\{24A_{2}\Big[\frac{7}{3}-(a_{2,1}+a_{2,2}+\frac{a_{2,3}}{3})\Big]+
24A_{3}\Big[\frac{7}{3}-(a_{3,1}+a_{3,2}+\frac{a_{3,3}}{3})\Big] {}\nonumber \\ &&+8A_{4}(1-a_{4,1}) \Big\}
\langle\sigma\rangle^{\frac{1+\beta}{\beta}}.
\end{eqnarray*}
In the above equation, there are seven unknown parameters which are needed to be obtained.

To this end, if we define more proper parameters in terms of the coefficients as, $z_{1}=a_{2,1}+a_{2,2}+
\frac{a_{2,3}}{3}$ and $z_{2}=a_{3,1}+a_{3,2}+\frac{a_{3,3}}{3}$. And assuming intuitively that $z_1$ and
$z_2$ have approximately equal values, one can define $a_{4,1}$ in terms of $z_{1}$ and $z_{2}$ as $a_{4,1}=
\frac{3}{7}\frac{z_{1}+z_{2}}{2}$. This final approximation can be elucidated if we consider Fig. 2. From this
figure one can see that $A_{4}$ assumes values a lot less than all the others, $A_{1}$, $A_{2}$ and $A_{3}$.
Meaning that any rough approximate values of $a_{4,1}$ can only change the final result very slightly (or even
the effect of it is unnoticeable). Thus the number of unknown parameters are reduced to two unknown, $z_{1}$ and $z_{2}$.

To obtain these two unknowns, one can exploit the physical properties of the average magnetization. For example,
if $K$ goes to infinity, $\langle\sigma\rangle$ must goes to one. Unfortunately, this property of average magnetization
can not produce a relation between $z_{1}$ and $z_{2}$.

Using the critical behavior of the average magnetization, one can use the property of $\langle\sigma\rangle$ at
the critical point: it assumes the value of zero if $K<K_{c}$, and it assumes values different from zero if $K>K_{c}$.

Using this final remark and assuming $z_{1}\simeq z_{2}$, and
denoting both of them with $z $, one can obtain the value of z as
$z=0.8238$ if the critical value of $K$ for body-centered cubic
lattice $K_{c}=0.1573$ is considered. After this short analysis,
we obtain the following relation for the average magnetization of
body centered cubic Ising lattice as,
\begin{equation}
\langle\sigma\rangle=\Big[\frac{1-8A_{1}-24(A_{2}+A_{3}+\frac{1}{7}A_{4})}{24(A_{2}+A_{3})(\frac{7}{3}-z)+
8A_{4}(1-\frac{3}{7}z)}\Big]^{\beta},
\end{equation}
where the value of $\beta$ can be taken as equal to $0.325$ for all the 3D Ising lattice.

We have plotted Fig. 3 for the obtained average magnetization expression presented with Eq. (3). In this figure,
the red data points indicates the simulation results of \cite{Hakan1} and dashed curve shows the result of Eq. (3).
\begin{figure*}[!hbt]
\begin{center}$
\begin{array}{cc}
\scalebox{1.1}{\includegraphics{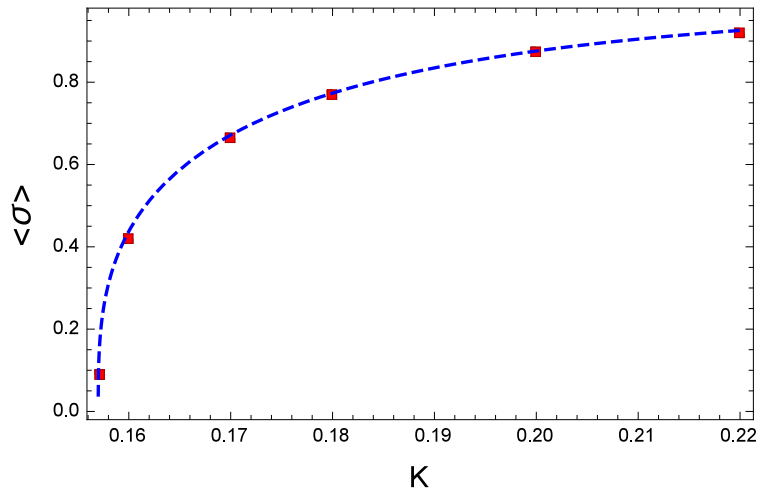}}
\end{array}$
\end{center}
\caption{The plot of the average magnetization of the body centered cubic Ising lattice with respect to $K$. }
\end{figure*}

As seen in Fig. (3), the agreement of the obtained expression for average magnetization of body-centered cubic
lattice in this paper and the simulation data is almost perfect. Even, it is almost impossible to see the differences
in this figure.

Therefore, we can claim that the heuristically obtained average magnetization relation is given by Eq. (3) is almost
exact. This means that the conjectures used here in this paper have been previously applied to the square, honeycomb,
triangular, and simple cubic lattices in references \cite{kaya21,kaya31} for the odd spins correlation functions are
relevant and also valuable.

\section{Conclusion and Discussion}
In this paper, the calculation of the average magnetization of the body-centered cubic Ising lattice was considered by
the method which was developed previously in a series of papers of us. Indeed, we have obtained an average magnetization
relation for this lattice structure, and we observed that the obtained expression is in good agreement with the
simulation data. Recalling, the almost exact agreement of the previously obtained result for honeycomb, square,
triangular, and simple cubic structure average magnetization relation and the average magnetization relation obtained
in this current paper for the body-centered cubic lattice.

We can claim that the conjectured mathematical forms for the odd spins correlation functions are relevant and also
important since they lead to an analytical calculation of the average magnetization of Ising lattices. Of course,
the method is inevitably heuristic. However, we think that any relevant analytical calculation method developed for
the calculation of the average magnetization expressions of Ising lattices can be considered valuable and helpful.

In other words, in this paper, we obtain the average magnetization relation of the body-centered cubic Ising lattice
with very simple and tractable mathematical and physical approaches. We think that this point is also important if
we recall the mathematical procedure applied to obtain the square lattice average magnetization by Yang.


\begin{thebibliography}{00}
\bibitem{Ising1}E. Ising, Contribution to the theory of ferromagnetism, Z. Phys. {\bf 31} (1925) 253-258.
\bibitem{Stanley1} H.E. Stanley, Introduction to Phase Transition and Critical Phenomena, Oxford University Press, New York, 1971.
\bibitem{Hu1} C.-K. Hu, Historical review on analytic, Monte Carlo, and renormalization group approaches to critical phenomena of some lattice models, Chin. J. Phys. {\bf 52} (2014) 1-76.
\bibitem{Zhu1} C.-P. Zhu, L.-T. Jia, L.-L. Sun, B.J. Kim, B.-H. Wang, C.-K. Hu, H.E. Stanley, Scaling relations and finite-size scaling in gravitationally correlated lattice percolation models, Chin. J. Phys. {\bf 64} (2020) 25-34.
\bibitem{Keskin1} N. \c{S}arl\i, M. Keskin, Effect of the distance range between the YBa-core and CuO-shell on the superconducting properties in the YBCO by an Ising model, Chin. J. Phys. {\bf 63} (2020) 375-381.
\bibitem{Aouini1} S. Aouini, A. Mhirech, A. Alaoui-Ismaili, L. Bahmad, Phase diagrams and magnetic properties of a double fullerene structure with core/shell, Chin. J. Phys. {\bf 59} (2019) 346-356.
\bibitem{Huang1} K. Huang, Statistical Mechanics, John Wiley and Sons, New York, 1987.
\bibitem{Baxter1} R.J. Baxter, Exactly Solved Model in Statistical Mechanics, Academic Press, London, 1982.
\bibitem{book1} S. Friedli, Y. Velenik, Statistical Physics of Lattice Systems, Cambridge University Press, 2017.
\bibitem{Beale1} R.K. Pathria, P.D. Beale, Statistical Mechanics, third ed.,  Elseveir, 2011, pp. 498.
\bibitem{McCOY1} B.M. McCoy, Advanced Statistical Mechanics, Oxford University Press, 2010.
\bibitem{Hasenbusch1}M. Hasenbusch, Monte Carlo studies of the three-dimensional Ising model in equilibrium, Int. J. Mod. Phys. C {\bf 12} (2001) 911-1009.
\bibitem{Jasch1} F. Jasch, H. Kleinert, Fast-convergent resummation algorithm and critical exponents of $\phi^4$ - theory in three dimensions, J. Math. Phys. {\bf 42} (2001) 52-73.
\bibitem{Gupta1}R. Gupta, P. Tamayo, Critical exponents of the 3-D Ising model, Int. J. Mod. Phys. C {\bf 7} (1996) 305-319.
\bibitem{Heringa1} H.W.J. Bl\"{o}te, J.R. Heringa, A. Hoogland, E.W. Meyer, T.S. Smit, Monte Carlo renormalization of the 3D Ising Model: Analyticity and convergence, Phys. Rev. Lett. {\bf 76} (1996) 2613-2616.
\bibitem{Salman1} Z. Salman, J. Adler, High and low temperature series estimates for the critical temperature of the 3D Ising model,  Int. J. Mod. Phys. C {\bf 9} (1998) 195-209.
\bibitem{Campostrini1} M. Campostrini, A. Pelissetto, P. Rossi, E. Vicari, 2002. 25th-order high-temperature expansion results for three-dimensional Ising-like systems on the simple-cubic lattice, Phys. Rev. E {\bf  65}, 066127.
\bibitem{Butera1} P. Butera, M. Comi, Extension to order $\beta^23$
 of the high-temperature expansions for the spin-$\frac{1}{2}$ Ising model on simple cubic and body-centered cubic lattices, Phys. Rev. B {\bf 62} (2000) 14837-14843.
\bibitem{Pelissetto1} A. Pelissetto, E. Vicari, Critical phenomena and renormalization-group theory, Phys. Rep. {\bf 368} (2002) 549-727.
\bibitem{Onsager1} L. Onsager, Crystal statistics. I. A two-dimensional model with an order-disorder transition, Phys. Rev. {\bf  65} (1944) 117-149.
\bibitem{Yang1} C.N. Yang, The spontaneous magnetization of a two-dimensional Ising model, Phys. Rev. {\bf 85} (1952) 808-816.
 \bibitem{Zhang1} Z.D. Zhang, Conjectures on the exact solution of three-dimensional (3D) simple orthorhombic Ising lattices,  Philos. Mag. {\bf 87} (2007) 5309-5419.
\bibitem{Wu41} F.Y. Wu, B.M. McCoy, M.E. Fisher, L. Chayes, Comment on a recent conjectured solution of the three-dimensional Ising model, Philos. Mag. {\bf 88} (2008) 3093-3095.
\bibitem{Perk51}J.H.H. Perk, Comment on 'Conjectures on exact solution of three-dimensional (3D) simple orthorhombic Ising lattices, Philos. Mag. {\bf  89} (2009) 761-764.
\bibitem{Yang831} C.N. Yang, Journey through statistical physics, Int. J. Mod. Phys. B {\bf 2} (1988) 1325-1329.
\bibitem{Perk11}J.H.H. Perk, Quadratic identities for Ising model correlations, Phys. Lett. A {\bf 79} (1980) 3-5.
\bibitem{Perk21}J.H.H. Perk, 2013. Comment on Mathematical structure of the three-dimensional (3D) Ising model, Chin. Phys. B  {\bf 22}, 080508.
\bibitem{Kramers1}H.A. Kramers, G.H. Wannier, Statistics of the two-dimensional ferromagnet. Part I, Phys. Rev. {\bf 60} (1941) 252-262.
\bibitem{Newell1} G.F. Newell, E.W. Montroll, On the theory of the Ising model of ferromagnetism, Rev. Mod. Phys. {\bf 25} (1953) 353-389.
\bibitem{Kaufman1} B. Kaufman, Crystal statistics. II. Partition function evaluated by spinor analysis, Phys. Rev. {\bf 76} (1949) 1232-1243.
\bibitem{Keh1}K.Y. Lin, Spontaneous magnetization of the Ising model, Chin. J. Phys. (Taipei) {\bf 30} (1992) 287-320.
\bibitem{Yang19521} C.N. Yang, The spontaneous magnetization of a two-dimensional Ising model, Phys. Rev. {\bf 85} (1952) 808-816.
\bibitem{201}I. Syozi, Transformation of Ising Models, in: C. Domb, M.S. Green (Eds.), Phase Transitions and Critical Phenomena Vol. 1: Exact  Results, Academic Press, New York, 1972, pp. 269-329.
\bibitem{261} M. Kac, J.C. Ward, A combinatorial solution of the two-dimensional Ising model, Phys. Rev. {\bf 88} (1952) 1332-1337.
\bibitem{271} R.B. Potts, J.C. Ward, The combinatrial method and the two-dimensional Ising model, Prog. Theor. Phys. {\bf 13} (1955) 38-46.
\bibitem{281} C.A. Hurst, H.S. Green, New solution of the Ising problem for a rectangular lattice, J. Chem. Phys. {\bf 33} (1960) 1059-1062.
\bibitem{331}E.W. Montroll, Lattice Statistics, in: E.F. Beckenbach, G. P\'{o}lya, Applied Combinatorial Mathematics Ch. 4, J. Wiley,  New York, 1964, pp. 96-143.
\bibitem{341} B.M. McCoy, T.T. Wu, The Two-Dimensional Ising Model, Harvard University Press, 1973.
\bibitem{Syozi1}I. Syozi, The statistics of honeycomb and triangular lattice. II, Prog. Theor. Phys. {\bf 5} (1950) 341-351.
\bibitem{Naya1}S. Naya, On the Spontaneous magnetizations of honeycomb and Kagomé Ising lattices, Prog. Theor. Phys. {\bf 11} (1954) 53-62.
\bibitem{Potts1} R.B. Potts, Spontaneous magnetization of a triangular Ising lattice, Phys. Rev. {\bf 88} (1952) 352.
\bibitem{1141} R.J. Baxter, Triplet order parameter of the triangular Ising model, J. Phys. A {\bf 8} (1975) 1797-1805.
\bibitem{Baxtery1}R.J. Baxter, T.C. Choy, Local three-spin correlations in the free-fermion and planar Ising models, Proc. R. Soc. Lond. A {\bf 423} (1989) 279-300.
\bibitem{Enting1} I.G. Enting, Triplet order parameters in triangular and honeycomb Ising models, J. Phys. A {\bf 10} (1977) 1737-1743.
\bibitem{1041} K.Y. Lin, M. Chen, Spontaneous magnetization and three-spin correlation of the Ising model on the isotropic checkerboard lattice, Chin. J. Phys. (Taipei) {\bf 27}, (1989) 79-89.
\bibitem{1051}  K.Y. Lin, Three-spin correlation of the Ising model on the generalized checkerboared lattice, J. Stat. Phys. {\bf 56} (1989) 631-643.
\bibitem{1061}  K.Y. Lin, Three-spin correlation of the Ising model on the anisotropic checkerboard lattice, Chin. J. Phys. {\bf 28} (1990) 159-172.
\bibitem{Tanaka1}J.H. Barry, C.H. Munera, T. Tanaka, Exact solutions for Ising model odd-number correlations on the honeycomb and triangular lattices, Physica A {\bf 113 } (1982) 367-387.
\bibitem{kaya21}T. Kaya, Relevant alternative analytic average magnetization calculation method for the square and the honeycomb Ising lattices, Chin. J. Phys. {\bf 77} (2022) 747-752.
\bibitem{kaya31}T. Kaya, Relevant spontaneous magnetization relations for the triangular and the cubic lattice Ising model, Chin. J. Phys. (2022) https://doi.org/10.1016/j.cjph.2022.03.043.
\bibitem{kaya1}T. Kaya, Exact three spin correlation function relations for the square and
the honeycomb Ising lattices, Chin. J. Phys. {\bf 66} (2020) 415-421.
\bibitem{Hakan1} P.H. Lundow, K. Markstr\"{o}m, A. Rosengren, The Ising model for the bcc, fcc and diamond
lattices: A comparison. Philos. Mag. {\bf 89} (2009) 2009-2042.
\end{thebibliography}
\end{document}